\documentclass[aps,twocolumn,showpacs]{revtex4}
\usepackage{graphicx}

\begin{document}


\title{The Dynamics of Hierarchical Evolution of Complex Networks}

\author{Matheus Palhares Viana}
\author{Luciano da Fontoura Costa}
\affiliation{
    Electronic address: luciano@if.sc.usp.br \\
    Instituto de F{\'i}sica de S{\~a}o Carlos, Universidade de S{\~a}o Paulo\\
    Av. Trabalhador S{\~a}o Carlense 400, Caixa Postal 369, \\
    CEP 13560-970, São Carlos, S{\~a}o Paulo, Brazil
    }

\date{1st May 2005}

\begin{abstract}
Introduced recently, the concept of hierarchical degree allows a more
complete characterization of the topological context of a node in a
complex network than the traditional node degree.  This article
presents analytical characterization and studies of the density of
hierarchical degrees in random and scale free networks.  The obtained
results allowed the identification of a hierarchy-dependent power law
for the degrees of nodes in random complex networks, with Poisson
density for the first hierarchical degree (obtained through master
equation approach).  Exact results were obtained for the second
hierarchical degree in scale free networks.

\end{abstract}

\maketitle

\section{\label{sec:level1}Introduction}

The characterization and analysis of complex networks~\cite{1,2}
involves the use of measurements capable of expressing important
properties of the network.  While traditional features such as the
node degree and clustering coefficient have been successfully used for
that purpose, such measurements are intrinsically local, in the sense
that they are affected only by the immediate neighborhood of each
reference node.  Among more global measurements, such as the network
diameter and the shortest path between any two network nodes, the
concept of \emph{hierarchical node degree} has been introduced
recently and considered for the characterization and analysis of the
connectivity of complex networks~\cite{4,5,6,7,8}.  Given a reference
node $i$, it is possible to define the $k-th$ relative hierarchical
level as corresponding to the network nodes which are exactly at
minimal distance of $k$ edges from node $i$.  The hierarchical node
degree at level $k$ corresponds to the number of edges between the
hierarchical levels $k$ and $k+1$, therefore defining a natural
hierarchical extension of the concept of node degree.  Note that the
hierarchical degree, starting at level $1$ (which includes the
immediate neighbors of $i$) tends first to increase with the
hierarchical level $k$ and then to decrease as a consequence of the
finite size of the network~\cite{9}.  Because the hierarchical node
degree is a global measurement (i.e. it considers, in hierarchical
fashion, all nodes in the network), it provides more information about
the connectivity of the reference node with the remainder of the
network.  For instance, the number of hierarchical levels which are
required in order to contain a percentage (e.g. 50\%) of the network
nodes, provides a valuable information about the accessibility of that
reference node.  The smaller this number of levels, the more connected
the reference node is with the rest of the network.  Several other
topological properties characterizing the context of the reference
node can be identified by using the hierarchical node
degree and other related features~\cite{7,8}.

An interesting question implied by the introduction of the concept of
hierarchical node degree regards the analytical characterization of
such features expected from random and scale free networks.  The
current work addresses such an issue by using mean field and the
master equation methodologies~\cite{3,10}.  It is shown that a power
law is obtained for the hierarchical degrees of nodes in random
networks, with Poisson density being obtained for the first
hierarchical degree.  We verified that the second order hierarchical
degree for scale free networks involves a logarithmic correction of
the first order degree studied in \cite{1,2,3}. In addition, we obtain
exact results for the second order hierarchical degree distributions
in scale free networks.

\subsection{\label{sec:level2}Hierarchy and hierarchical degree}

A set of hierarchy $h$ is defined as that containing all the
vertices which are located at a distance of $h$ connections from a
reference vertex. The hierarchical degree, or superior order
degree, is the measurement of the number of connections in a set
with a determined value of hierarchy. For $h=1$, we have the first
order degree which is well known in the literature \cite{1,2,3}.
In this article we are intended to study the behavior of graphs
where $h\geq2$. We will denote $k^{(h)}_i$ as the degree of order
$h$ of the reference vertex $i$; and the set of hierarchy $h$ of
the reference vertex $i$ will be denoted as $\{
i^{(h)}_1,i^{(h)}_2,i^{(h)}_3,...,i^{(h)}_{k^{(h-1)}_i } \}$.  We
will also denote $N^{(h)}_x$ as the number of vertices having
hierachical degree $h$ equal to $x$, and $N_{a,b,...,n}$ will
denote the number of vertices presenting the first order degree as
$a$, the second order degree as $b$ and so on.

\section{\label{sec:level3}Degree Evolution of a Vertex}

Mean field theory is widely applied in complex networks research
\cite{1,2,3}. Through this boarding, the degree evolution of the
particular vertice is proportional its connection probability
(\textit{kernel function} - $P_J(t)$). If this probability has a
complex form, the equations become non-trivial.

Generally, the hierarchical degree $h$ depends on the probability of
connection of the vertices belonging to this set of hierarchy. Thus, we
can write:

\begin{equation}
\frac{\partial k^{(h)}_j(t)}{\partial t} = \sum_{i=1}^{k^{(h-1)}_i}P_{j^{(h)}_i}+\delta_{1h}P_j(t)
\end{equation}

In the following we consider random graphs where $P_j(t)$ is the
same for all vertices, as well as scale-free networks proposed by
Barab{\'a}si-Albert \cite{1} where $P_j(t)\propto k^{(1)}_j$.

\section{\label{sec:level4} Distribution of the Second Order Connectivity Degree}

In this section we will study the connectivity degree distribution,
which is an intrinsic characteristic of the complex networks and is
capable of distinguishing all types of graphs. Henceforth, for
simplicity's sake, the probability of vertex $j$ to make a connection
will be represented as ${\cal P}_{k^{(1)}_j}(t)$ and not $P_j(t)$.

\subsection{\label{sec:level15}Initial States and Structures}

\begin{figure*}
    \includegraphics[scale=0.3]{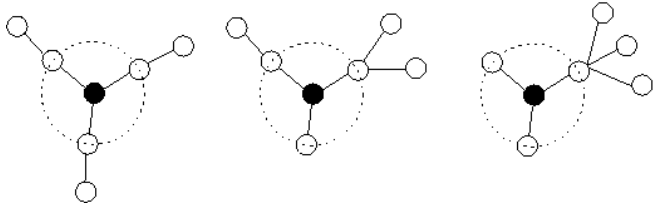}
    \caption{Possible instances of the structure $(3,3)$.}
    \label{estr33}
\end{figure*}

A vertex with first order degree $k^{(1)}$ and second order degree
$k^{(2)}$ defines one or more structures which are topologically
identic and are denoted by $(k^{(1)},k^{(2)})$. For example,
Figure~\ref{estr33} shows all possible structures $(3,3)$.

An structure has one or more initial states. An structure $(x_1,x_2)$
is the initial state of another structure $(y_1,y_2)$ if the following
condition is met:

\begin{equation}
    (x_1+\delta_1,x_2+\delta_2) \rightarrow (y_1,y_2)
\end{equation}

where:

\begin{equation}
    \delta_1+\delta_2 = 1
\end{equation}

Therefore, two processes exist which take one structure into
the other:

\begin{eqnarray}
    (x_1+1,y_2) & \rightarrow & (y_1,y_2)\\
    (y_1,x_2+1) & \rightarrow & (y_1,y_2)
\end{eqnarray}

\begin{figure*}
    \includegraphics[scale=0.4]{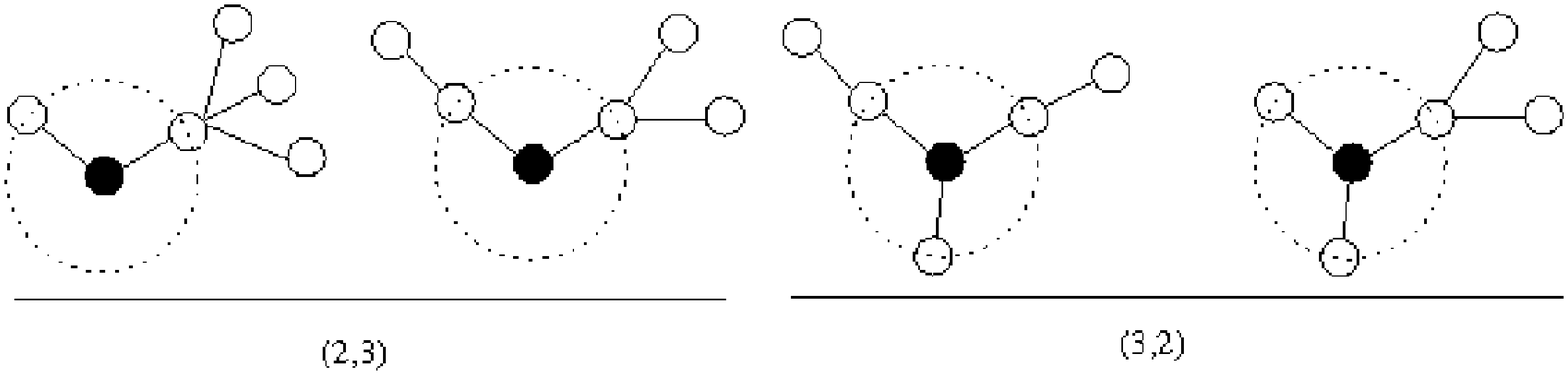 }
    \caption{Initial states of structure $(3,3)$.}
    \label{ini_str33}
\end{figure*}

The initial states of structure $(3,3)$, given by $(2,3)$ and $(3,2)$
are represented in Figure~\ref{ini_str33}.

\subsection{\label{sec:level16}Describing the Processes through the Master Equation}

The probability of transition of the initial states in a structure
give us the production rate of that structure. For example, let us
study the formation of structures $(k^{(1)},k^{(2)})$. As we know,
their initial states are: $(k^{(1)}-1,k^{(2)})$ and
$(k^{(1)},k^{(2)}-1)$.  Then, the transition probabilities are given
by:

\begin{eqnarray*}
    P_{ (k^{(1)}-1,k^{(2)})\rightarrow(k^{(1)},k^{(2)}) } &=& {\cal P}_{k^{(1)}-1}\\
    P_{ (k^{(1)},k^{(2)}-1)\rightarrow(k^{(1)},k^{(2)}) } &=& {\cal P}_{k^{(1)}+k^{(2)}-1}
\end{eqnarray*}

Therefore, the formation rate of the structure $(k^{(1)},k^{(2)})$ are:

\begin{eqnarray*}
    N_{+k^{(1)}k^{(2)}} & = & {\cal P}_{k^{(1)}-1}N_{k^{(1)}-1k^{(2)}} + {\cal P}_{k^{(1)}+k^{(2)}-1}N_{k^{(1)}k^{(2)}-1}\\
    & + & \delta_{1k^{(1)}}{\cal P}_{k^{(2)}}N^{(1)}_{k^{(2)}}
\end{eqnarray*}

The transition probabilities of structure $(k^{(1)},k^{(2)})$ is:

\begin{eqnarray*}
    P_{ (k^{(1)},k^{(2)})\rightarrow(k^{(1)}+1,k^{(2)}) } &=& {\cal P}_{k^{(1)}}\\
    P_{ (k^{(1)},k^{(2)})\rightarrow(k^{(1)},k^{(2)}+1) } &=& {\cal P}_{k^{(1)}+k^{(2)}}
\end{eqnarray*}

Therefore, the vanishing rate of the structure $(k^{(1)},k^{(2)})$ is:

\begin{eqnarray*}
    N_{-k^{(1)}k^{(2)}} & = & [{\cal P}_{k^{(1)}}+{\cal P}_{k^{(1)}+k^{(2)}}]N_{k^{(1)}k^{(2)}}
\end{eqnarray*}

The master equation governing the evolution of $N_{k^{(1)}k^{(2)}}$ is
as follows:

\begin{eqnarray}
\nonumber{} \frac{\partial N_{k^{(1)}k^{(2)}}}{\partial t} & = &  {\cal P}_{k^{(1)}-1}N_{k^{(1)}-1k^{(2)}} + {\cal              P}_{k^{(1)}+k^{(2)}-1}N_{k^{(1)}k^{(2)}-1} \\
\nonumber{} & - & [{\cal P}_{k^{(1)}}+{\cal P}_{k^{(1)}+k^{(2)}}]N_{k^{(1)}k^{(2)}}\\
    & + & \delta_{1k^{(1)}}{\cal P}_{k^{(2)}}N^{(1)}_{k^{(2)}}
\end{eqnarray}

In the following we will show the solution of the equation above for random
graphs as well as for scale-free networks.

\section{\label{sec:level17}Applications}

\subsection{\label{sec:level18}Random Graphs}

In this work we use non-static version of non-directed random
graphs. Instead of generating a network from a symmetric random
matrix, we define the following procedure: Beginning with ${\cal
N}$ disconnected points, we select two vertices in the network at
a time and, with probability $p$, we establish a connection
between them. Connections linking a vertex to itself are not
allowed, that is to say, the two selected vertices must be
different. The probability of a vertex $j$ to be selected at a
time $t$ is given by:

\begin{equation}
    P_j(t) = \frac{p}{{\cal N}}+\frac{p}{{\cal N}-1} = \frac{(2{\cal N}-1)}{{\cal N}({\cal N}-1)}p
\end{equation}

\subsubsection{\label{sec:level19}Degree Evolution of a Vertex}

Since $P_j(t)$ is constant, we will denote it as $\pi$. Therefore:

\begin{equation}
    \pi = P_j(t)
\end{equation}

By substituting Eq.(8) in Eq.(1), we have:

\begin{equation}
    \frac{\partial k^{(h)}_j(t)}{\partial t} = \pi \left( k^{(h-1)}_i+\delta_{1h} \right)
\end{equation}

Assuming that $k^{(h)}_j=g_ht^h$, we find $g_h = \pi^h/h!$ and

\begin{eqnarray}
    k^{(h)}_j(t) & = & \frac{(\pi t)^h}{h!}
\end{eqnarray}

\subsubsection{\label{sec:level110}First Order Degree Distribution}

The variation of the number of vertices with first order degree given as
$k^{(1)}$ can be described through the master equation:

\begin{eqnarray*}
\frac{\partial N^{(1)}_{k^{(1)}}}{\partial t} & = & \pi[N^{(1)}_{k^{(1)}-1}-N^{(1)}_{k^{(1)}}]
\end{eqnarray*}

As is known, $N^{(1)}_0(t=0)={\cal N}$ and $N_j(t=0)=0, \forall
j$. Solving this equation for $k^{(1)}=0$, we have:

\begin{eqnarray*}
    N^{(1)}_{0}(t) & = & {\cal N}e^{-\pi t}
\end{eqnarray*}

And for $k^{(1)}=1$, we have:

\begin{eqnarray*}
    N^{(1)}_1(t) & = & \pi{\cal N}te^{-\pi t} = \pi tN^{(1)}_0(t)
\end{eqnarray*}

For other values of $k^{(1)}$, we conclude that the solution
of the master equation obeys the following formation rule:

\begin{eqnarray*}
    N^{(1)}_{k^{(1)}}(t) & = & \frac{\pi t}{k^{(1)}}N^{(1)}_{k^{(1)}-1}(t)
\end{eqnarray*}

The solution for this recursion is:

\begin{eqnarray*}
    N^{(1)}_{k^{(1)}}(t) & = & {\cal N}\frac{(\pi t)^{k^{(1)}}e^{-\pi t}}{k^{(1)}!}
\end{eqnarray*}

The distribution of the first order degree, given by
 $P^{(1)}_{k^{(1)}}(t)=N^{(1)}_{k^{(1)}}(t)/{\cal N}$ is:

\begin{eqnarray}
    P^{(1)}_{k^{(1)}}(t) & = & \frac{(\pi t)^{k^{(1)}}e^{-\pi t}}{k^{(1)}!}
\end{eqnarray}

\begin{figure*}
    \includegraphics[scale=0.6]{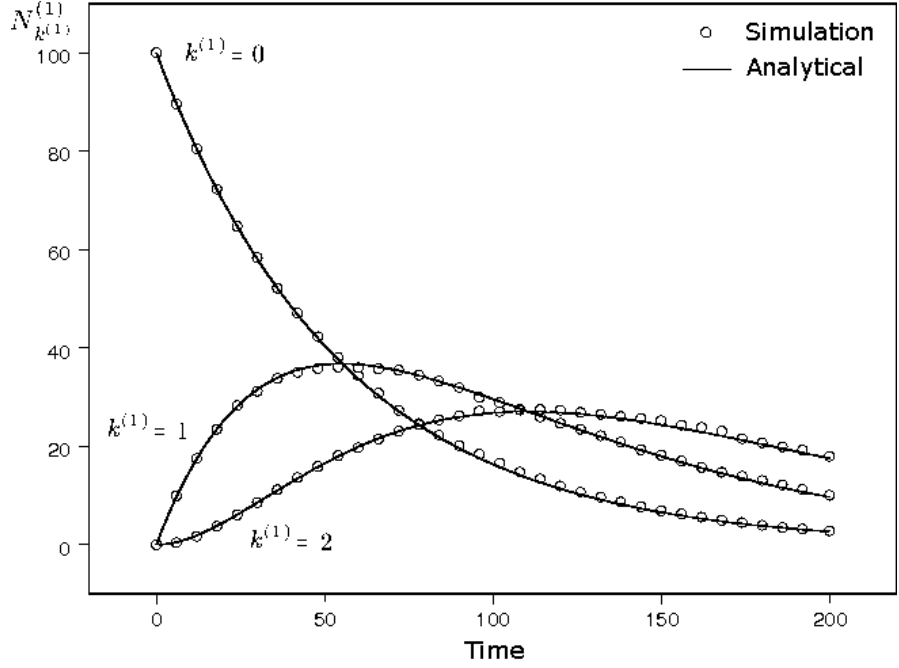}
    \caption{Evolution of the number of vertices with first order degree
    $k^{(1)}=\{0,1,2\}$ in a random graph with ${\cal N}=100$ and $p=0.9$.}
    \label{random}
\end{figure*}

This is the Poisson distribution with mean value $\pi
t$. Figure~\ref{random} shows the time evolution of the number of vertices with
$k^{(1)}={0,1,2}$ in a random graph with ${\cal N}=100$ and
$p=0.9$.

\subsection{\label{sec:level112}Scale-free Networks}

The concept of scale free networks has been introduced by
Barab{\'a}si and Albert and refers to graphs whose first order
degree distribution follows a power law \cite{3}:
$N_{k^{(1)}}=A[k^{(1)}]^{-\nu}$.  Most scale free networks found
in nature have $\nu$ in the interval between 2 and 3. The
probability of connection in this model is given by:

\begin{eqnarray}
    P_j(t) & = & \frac{k^{(1)}_j}{ \sum_{i=1}^{t}k^{(1)}_i } = \frac{k^{(1)}_j}{2t}
\end{eqnarray}

Also:

\begin{eqnarray}
    {\cal P}_{k^{(h)}}(t) & = & \frac{k^{(h-1)}+k^{(h)}}{ \sum_{i=1}^{t}k^{(1)}_i } = \frac{k^{(h-1)}+k^{(h)}}{2t}
\end{eqnarray}

\subsubsection{\label{sec:level113}Evolution of Vertex Degree}

Substituting Eq.(12) in Eq.(1), we have:

\begin{equation}
    \frac{\partial k^{(h)}_j(t)}{\partial t} = \frac{1}{2t}\sum_{i=1}^{k^{(h-1)}_j}k^{(1)}_{j^{(h)}_i}+\delta_{1h}\frac{k^{(1)}_j}{2t}
\end{equation}

>From the definition of hierarchical degree, we have:

\begin{equation}
    k^{(h)}_j + k^{(h-1)}_j = \delta_{1h}k^{(1)}_j + \sum_{i=1}^{k^{(h-1)}_j}k^{(1)}_{j^{(h)}_i}
\end{equation}

Therefore:

\begin{equation}
    \frac{\partial k^{(h)}_j(t)}{\partial t} = \frac{1}{2t} \left( k^{(h)}_j + k^{(h-1)}_j \right)
\end{equation}

For $h=1$ and the following condition: $k^{(1)}_j(t=t_j) = 1$ we
obtain the solution given by Barab{\'a}si-Albert in \cite{3}:

\begin{equation}
    k^{(1)}_j(t) = \sqrt{\frac{t}{t_j}}
\end{equation}

For $h=2$, we use the integrating factor $t^{-3/2}$ and we obtain:

\begin{eqnarray*}
    k^{(2)}_j(t) & = & \left[\beta_2+\ln{\sqrt{t}}\right]k^{(1)}_j(t)
\end{eqnarray*}

If we consider that, in instant where the new vertice is enclosed to network its average second order degree is given by $\kappa^{(2)}_j$, then: $k^{(2)}_j(t=t_j)=\kappa^{(2)}_j$. From this condition, we have:

\begin{eqnarray*}
    k^{(2)}_j(t) & = & \left[\kappa^{(2)}_j+\ln{k^{(1)}_j}\right]k^{(1)}_j(t)
\end{eqnarray*}

\begin{figure*}
    \includegraphics[scale=0.5]{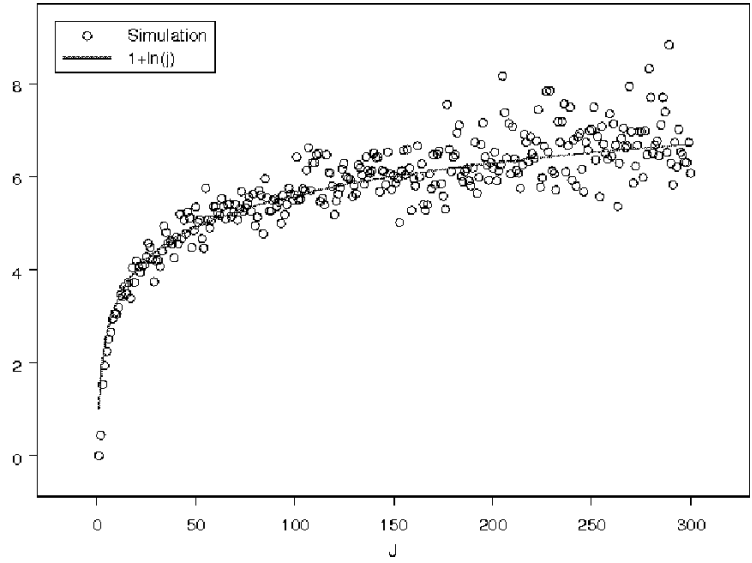}
    \caption{Values of $\kappa^{(2)}_j$ in terms of $j$ obtained through
    numerical simulations are adjusted by the curve $k^{(2)}_j=1+\ln{j}$.}
    \label{initial}
\end{figure*}

\begin{figure*}
    \includegraphics[scale=0.5]{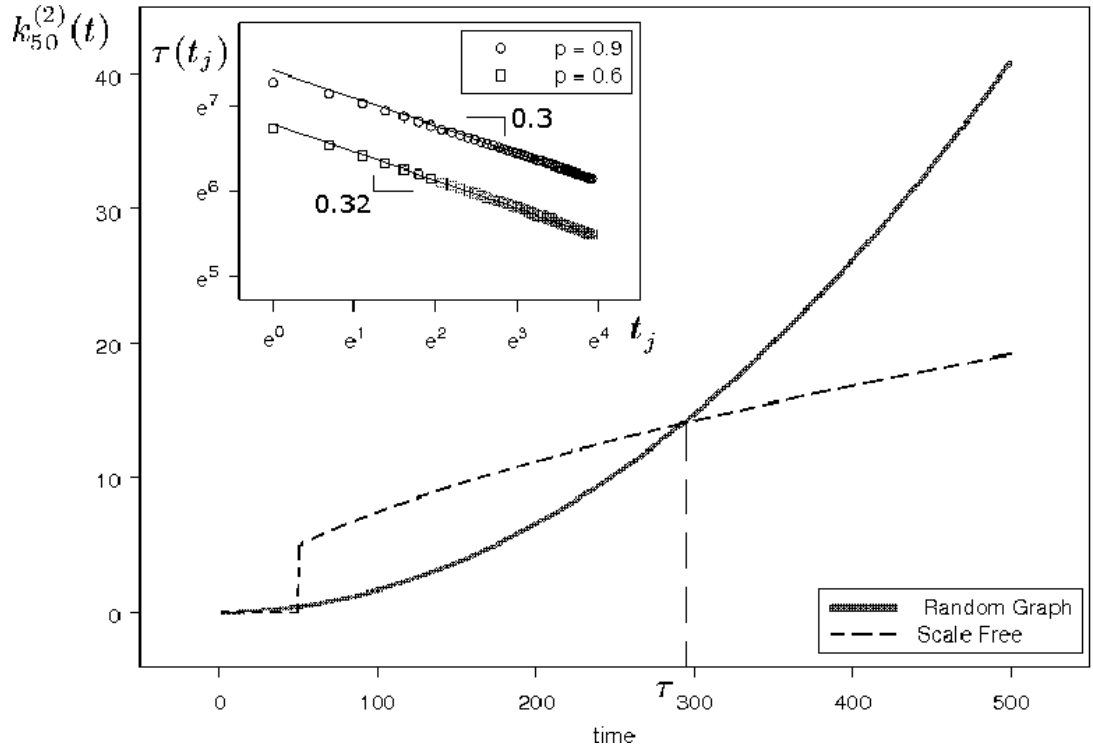}
    \caption{Evolution of $k_{50}^{(2)}(t)$ for a random
    graph with size ${\cal N}=100$ and connection probability $p=0.9$ and
    Scale-Free network. The curves intercept at $t=\tau$,
    indicating that at this growth stage vertex $j$ in the scale free
    network has the same second order degree as vertex $j$ in the random
    network. The log-log plot in the inset shows how $\tau(t_j)$ depends
    of $t_j$, defining power laws $\tau(t_j)\propto
    t_j^{-\alpha}$.~\label{fig:comp}}
    \label{k2}
\end{figure*}

Through numerical simulations, we can determine the value of
$\kappa^{(2)}_j$ in terms of $j$. From Fig(4) we note that
$\kappa^{(2)}_j = 1+\ln{j}$. Thus, the solution for the second
order degree variation is given by:

\begin{eqnarray}
    k^{(2)}_j(t) & = & \left[1+\ln{t_jk^{(1)}_j}\right]k^{(1)}_j(t)
\end{eqnarray}

Eq.(15) is not simple to be solved but, using the same integrating factor as above,
we can obtain a solution as an integral recursive equation, as follows:

\begin{equation}
    k^{(h)}_j(t) =  \beta_h k^{(1)}_j(t) + \frac{1}{2}\sqrt{t}\int{k^{(h-1)}_j t^{-\frac{3}{2}}dt }
\end{equation}

\subsubsection{\label{sec:level114}First Order Degree Distribution}

In \cite{10}, Krapvisky establishes the following master equation
which governs the evolution of $N^{(1)}_{k^{(1)}}$, that is, the
average number of vertices with first order degree $k^{(1)}$:

\begin{equation}
    \frac{\partial N^{(1)}_{k^{(1)}}}{\partial t}=\frac{1}{M(t)}\left[(k^{(1)}-1)N^{(1)}_{k^{(1)}-1}-k^{(1)}N^{(1)}_{k^{(1)}}\right]+\delta_{1k^{(1)}}
\end{equation}

The first term takes into account the probability that the new
vertex connects to one of the first order degree equal to
$k^{(1)}-1$.  The second term expresses the probability of
connection between a vertex with $k^{(1)}$.  The delta function
accounts for the fact that every added vertex has degree $1$.  By
making the normalization $M(t)=2t$, the solution of Eq.(15) is
given as:

\begin{equation}
N^{(1)}_{k^{(1)}}(t)=\frac{4t}{k^{(1)}(k^{(1)}+1)(k^{(1)}+2)}
\end{equation}

The distribution of the first order degree, i.e.
$P^{(1)}_{k^{(1)}}(t)=N^{(1)}_{k^{(1)}}(t)/N_T$, is given as:

\begin{equation}
    P^{(1)}_{k^{(1)}}(t)=\frac{4}{k^{(1)}(k^{(1)}+1)(k^{(1)}+2)}
\end{equation}

Figure~\ref{k2} shows the evolution of $k_{50}^{(2)}(t)$ for a
random graph with $N=100$ nodes and connection probability $p=0.9$ as well as the same measurement for a scale free network.  An intersection point can be observed at $t=\tau$, meaning
that at this growth stage the vertex $j$ in the random and scale free
networks have the same second order degree.  For values of $t$ larger
than $\tau$, the second order degree becomes larger for the random
network than the scale free counterpart.  This indicates that the
second order degree in a scale free network tends to grow slower than
in a random network.  The log-log diagram in the inset, shows
$\tau(t_j)$ in terms of $t_j$ for connection probabilities $p=0.9$ and $p=0.6$, which yields power laws $\tau(t_j)\propto
t_j^{-\alpha}$.

\subsubsection{\label{sec:level115}Second Order Degree Distribution}

We now consider the number of vertices with first order degree equal
to $k^{(1)}$ and second order degree equal to $k^{(2)}$, which will be
represented as $N_{k^{(1)}k^{(2)}}$.  The obtention of this quantity
involves the solution of Eq.(6). Assuming a linear solution of the type: $N_{k^{(1)}k^{(2)}} = n_{k^{(1)}k^{(2)}}t$, we have:

\begin{eqnarray*}
    n_{k^{(1)}k^{(2)}} & = &  \frac{ {\cal P}_{k^{(1)}-1}n_{k^{(1)}-1k^{(2)}} + {\cal P}_{k^{(1)}+k^{(2)}-1}n_{k^{(1)}k^{(2)}-1} }{ {\cal       P}_{k^{(1)}}+{\cal P}_{k^{(1)}+k^{(2)}}+1/t }\\
    & + & \frac{ \delta_{1k^{(1)}}{\cal P}_{k^{(2)}}n_{k^{(2)}} }{ {\cal P}_1+{\cal P}_{1+k^{(2)}}+1/t }
\end{eqnarray*}

Substituting Eq.(13) into above expression:

\begin{eqnarray*}
    n_{k^{(1)}k^{(2)}} & = &  \left[ \frac{ (k^{(1)}-1)n_{k^{(1)}-1k^{(2)}} }{ 2k^{(1)}+k^{(2)}+2 } \right]\\
    & + & \left( \frac {k^{(1)}+k^{(2)}-1 }{ 2k^{(1)}+k^{(2)}+2 }\right)n_{k^{(1)}k^{(2)}-1}\\
    & + & \left( \frac{ k^{(2)}n_{k^{(2)}} }{ 4+k^{(2)} } \right)\delta_{1k^{(1)}}
\end{eqnarray*}

For simplicity's sake, we make

\begin{eqnarray}
    u & = & k^{(2)}\\
    A_u & = & n_{k^{(1)}u}\\
    F_u & = & \left[ \frac{ (k^{(1)}-1)n_{k^{(1)}-1u} }{ 2k^{(1)}+u+2 } \right] +\left( \frac{ un_{u} }{ 4+u } \right)\delta_{1k^{(1)}}\\
    G_u & = & \frac{ k^{(1)}+u-1 }{ 2k^{(1)}+u+2 }
\end{eqnarray}
\\

It follows that

\begin{equation}
    A_u = F_u+G_uA_{u-1}
\end{equation}

In addition, we have to enforce that

\begin{equation}
    A_1 = F_1,\;\;\;\;\;\;n_{k^{(1)}0}=0
\end{equation}

Developing the recurrence, we obtain

\begin{eqnarray*}
    A_u &=&  F_u + G_u (F_{u-1}+G_{u-1}A_{u-2})\\
    A_u &=&  F_u + G_u F_{u-1}+G_uG_{u-1}(F_{u-2}+G{u-2}A_{u-3})...
\end{eqnarray*}

\begin{eqnarray}
    A_u & = &  \frac{1}{G_{u+1}}\sum_{f=1}^{u}F_f\prod_{g=f}^{u}G_{g+1}
\end{eqnarray}

Expressed in terms of the original variables, it follows that

\begin{widetext}
    \begin{eqnarray*}
        n_{k^{(1)}k^{(2)}}
        & = &  \left(\frac{ 2k^{(1)}+k^{(2)}+3 }{ k^{(1)}+k^{(2)} }\right)
        \sum_{f=1}^{k^{(2)}}\left[ \frac{ (k^{(1)}-1)n_{k^{(1)}-1f} }{ 2k^{(1)}+f+2 } \right]
        \prod_{g=f}^{k^{(2)}}\left(\frac{ k^{(1)}+g }{ 2k^{(1)}+g+3 }\right)\\
        & + &  \frac{\Gamma(k^{(2)}+1)}{\Gamma(k^{(2)}+5)}
        \sum_{f=1}^{k^{(2)}}f(f+1)(f+2)(f+3)n_{f}
    \end{eqnarray*}
\end{widetext}

\begin{figure*}
    \includegraphics[height=6cm,width=14cm]{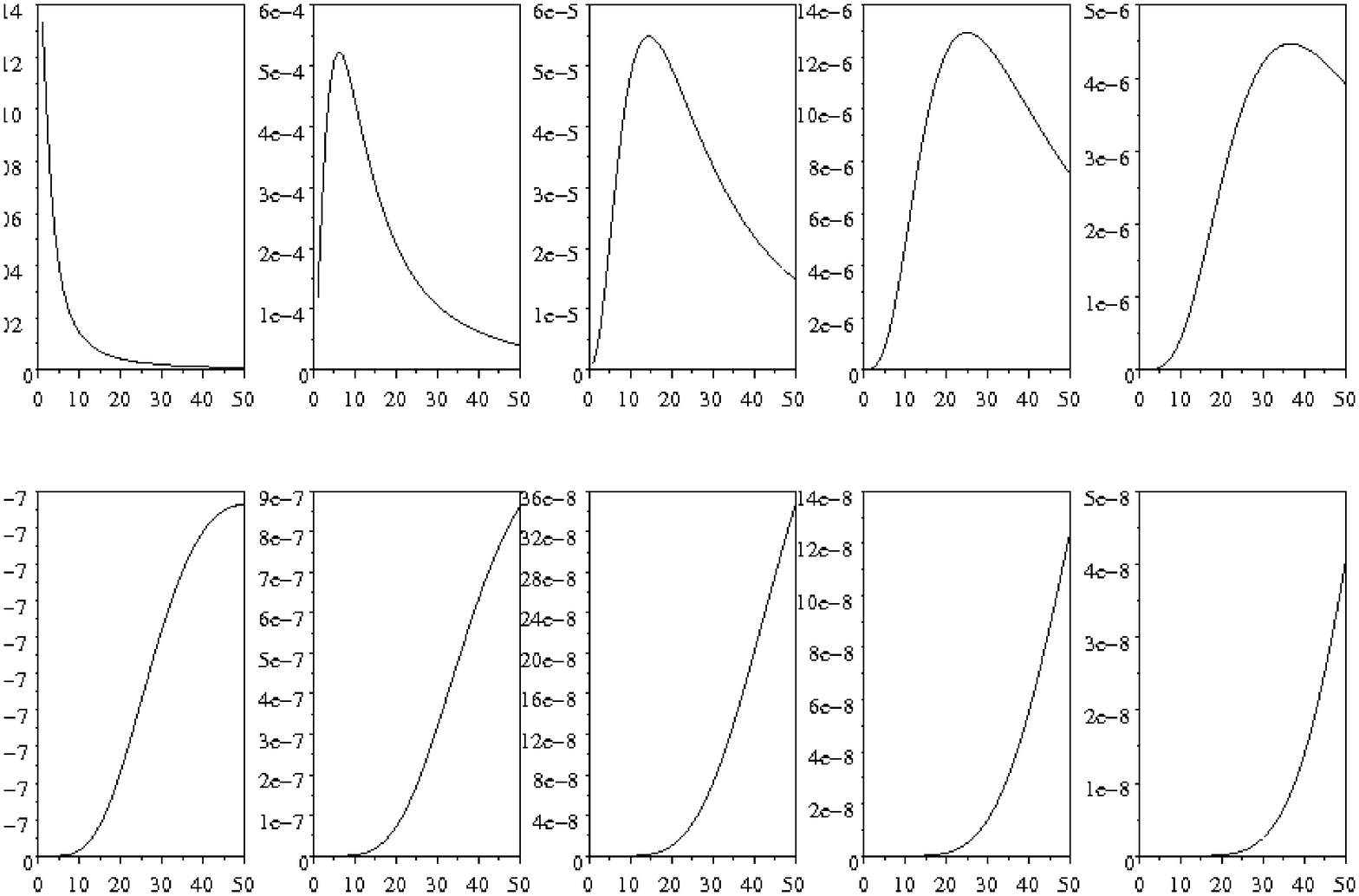}
    \caption{Values of $n_{k^{(1)}k^{(2)}}$ for $k^{(1)}=[1,5,10,15,20,25,30,35,40,45]$ and $k^{(2)}$ in the range 0,50.}
    \label{all}
\end{figure*}

\begin{figure*}
    \begin{center}
        \includegraphics[scale=0.5]{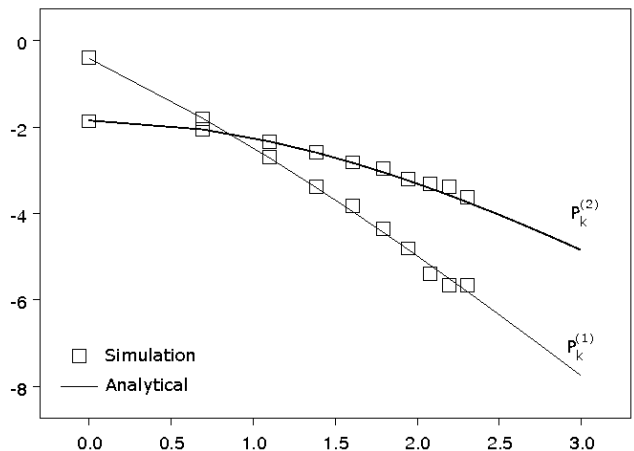}
        \caption{Log-log distribution of $P^{(1)}_k$ and $P^{(2)}_k$ for Barab{\'a}si-Albert model of Scale Free Network.}
        \label{nk1nk2}
    \end{center}
\end{figure*}

By Eq.(21) we know the value of $n_f$ ($n_f=N_f(t)/t$). Substituting in the above expression:

\begin{widetext}
    \begin{eqnarray}
        n_{k^{(1)}k^{(2)}} =  \left(\frac{ 2k^{(1)}+k^{(2)}+3 }{ k^{(1)}+k^{(2)} }\right)
        \sum_{f=1}^{k^{(2)}}\left[ \frac{ (k^{(1)}-1)n_{k^{(1)}-1f} }{ 2k^{(1)}+f+2 } \right]
        \prod_{g=f}^{k^{(2)}}\left(\frac{ k^{(1)}+g }{ 2k^{(1)}+g+3 }\right)+
        \frac{2k^{(2)}(k^{(2)}+7)}{(k^{(2)}+1)(k^{(2)}+2)(k^{(2)}+3)(k^{(2)}+4)}
    \end{eqnarray}
\end{widetext}

When $k^{(1)}=1$, we have that

\begin{eqnarray}
    n_{1k^{(2)}} & = &  \frac{2k^{(2)}(k^{(2)}+7)}{(k^{(2)}+1)(k^{(2)}+2)(k^{(2)}+3)(k^{(2)}+4)}
\end{eqnarray}

This result is identical to the expression obtained by Krapivsky in \cite{10} for correlation degree. Developing the first part of Eq. (30):

\begin{widetext}
    \begin{eqnarray}
        n_{k^{(1)}k^{(2)}} &=&  n_{1k^{(2)}}+
        \left(\frac{  k^{(1)}-1 }{  2k^{(1)}+k^{(2)}+2  }\right)
        \sum_{f=0}^{k^{(2)}-1}\left[ \frac{ \Gamma(2k^{(1)}+f+2)  }{  \Gamma(k^{(1)}+f)  } \right]n_{k^{(1)}-1f}
    \end{eqnarray}
\end{widetext}

Figure~\ref{all} shows the curves obtained for $n_{k^{(2)}k^{(2)}}$ for several
values of $k^{(1)}$ and $k^{(2)}$ varying from 0 to 50.

The total number of vertices with second order degree equal to
$k^{(2)}$ is given as:

\begin{equation}
    N^{(2)}_{k^{(2)}}(t)=t\sum_{k^{(1)}=1}^{t}n_{k^{(1)}k^{(2)}}
\end{equation}

Figure~\ref{nk1nk2} shows a log-log graph of the distribution $N^{(1)}_{k^{(1)}}$ (given
by Eq.(21)) and $N^{(2)}_{k^{(2)}}$ (obtained from Eq.(33)). Note that the
initial values in this distribution deviate from the linear relation
typically observed for scale free networks. The second order distribution, given by
$P^{(2)}_{k^{(2)}}(t)=N^{(2)}_{k^{(1)}}(t)/N_T$.

\section{Concluding Remarks}

This work addressed the analytical characterization of
hierarchical degrees of random and complex networks.  In the case
of random networks, we have shown that the hierarchical degree
follows a power law, i.e.  $k^{(h)}(t)\propto t^h$.  By using the
master equation approach, it has also been shown that the first
order degree obeys a Poisson distribution.  Unlike the first order
degree, the evolution of the second order hierarchical degree can
not be easily described by a master equation as Eq.(6).  This is a
consequence of the fact that the adopted approach of uniting two
connected components from the network into a single component
implies several combinations of effects and respective terms to be
incorporated into the master equation.  In addition, we verified
that the second order degree tends to grow slower in scale free
networks than in random networks, with the position where these
two values become equal following a power law.

In the case of scale free models, namely the Barab\'asi-Albert
network, it has been verified that the second order hierarchical
degree, for a particular node, can be expressed in terms of a
logarithmic correction of the first order degree.  Exact results have
been obtained for the second order hierarchical degree.  We observe
that the generalization of such an approach to higher hierarchical
levels becomes substantially more complex because of the need to
consider all recursions up to level $h-1$.

Possible continuations of the reported developments include the
extension of the analytical expressions of hierarchical node degree to
higher levels, as well as the derivation of analytical expression for
other hierarchical features such as the \emph{hierarchical clustering
coefficient} and \emph{hierarchical number of
nodes}~\cite{4}.

{\bf Acknowledgments}
Luciano da F. Costa thanks HFSP RGP39/2002, FAPESP (proc. 99/12765-2)
and CNPq (proc. 3082231/03-1) for financial support.

\end{document}